\documentstyle[12pt]{article}

\let\ni=\noindent

\begin{document}

\baselineskip 0.85cm

\pagestyle {empty}

~~~

\vspace{1.5cm}

\newcommand{\rt}{(\vec{r},t)}

\newcommand{\vr}{\vec{r}}

\newcommand{\lG}{\lambda_\Gamma}

{\large \centerline {\bf Hypothetic time---temperature duality as a hint}}
{\large \centerline {\bf for modifications in quantum dynamics?}}

\vspace{0.6cm}

{\centerline {\sc Wojciech Kr\'{o}likowski}}

\vspace{0.6cm}

{\centerline {\it Institute of Theoretical Physics, Warsaw University}}

{\centerline {\it Ho\.{z}a 69,~~PL--00--681 Warszawa, ~Poland}}

\vspace{0.6cm}

{\centerline {\bf Abstract}}

\vspace{0.2cm}

 The well--known formal analogy between time and absolute temperature, 
existing on the quantum level, is considered as a profound duality relationship
requiring some modifications in the conventional quantum dynamics. They consist
of tiny deviations from uniform time run in the physical spacetime, as well as 
of tiny deviations from unitary time evolution characteristic for the con%
ventional quantum theory. The first deviations are conjectured to be produced 
by local changes of total average particle number. Then, they imply the second 
deviations exerting in turn influence upon this particle number. Two examples 
of the second deviations are described: hypothetic tiny violation of optical 
theorem for particle scattering, and hypothetic slow variation of the average 
number of probe particles contained in a sample situated in proximity of a big 
acceler\-ator (producing abundantly particles on a target).   
   
\vspace{0.5cm} 

\ni PACS numbers:  11.10.Lm , 11.90.+t , 12.90.+b .  

\vspace{-20cm}
\begin{flushright}
{\bf IFT--98/24}
\end{flushright}
\vspace{20cm}
\vspace{2.0cm}


\vfill\eject

\pagestyle {plain}

\setcounter{page}{1}

\vspace{0.3cm}

\ni {\bf 1. Introduction}

\vspace{0.3cm}

 As is well known, in the quantum theory there exists a formal analogy [1] 
between the unitary time evolution described by the operator $\exp(-iHt/\hbar)$
and the thermal equilibrium connected with the operator  $\exp(-H/kT)$ (so, the
unitary time evolution may be also referred to as the "quantum dynamical 
equilibrium"). This may  raise a natural question, as to whether the implied 
correspondence between time and absolute temperature,

\begin{equation}
it/\hbar \leftrightarrow 1/kT \; ,
\end{equation}

\ni is only a formal analogy, appearing more or less accidentally in the 
structure of quantum theory, or is rather a profound relationship expressing 
a fundamental duality of time and absolute temperature (when $ H $ is time--%
independent, what is the generic situation from the theoretical viewpoint).%
{\footnote{It should be emphasized that the correspondence (1) implies on the 
quantum level a thermodynamic--like character of time $ t $ in contrast to its 
more familiar status on the classical level. Such a statistical--like character
of time is mathematically independent of (but physically consistent with) the 
fundamental probabilistic interpretation of the wave function in conventional 
quantum theory and, naturally, the probabilistic interpretation of the density
matrix in conventional statistical quantum theory, built up on top of the
former.}} In the second case, a fascinating possibility is open [2]: for 
quantum dynamical systems in some circumstances there might appear tiny but 
measurable deviations from the unitary time evolution, in analogy with the 
familiar deviations from the thermal equilibrium.{\footnote{Then, both notions 
of time and temperature lose their conventional meaning defined by the 
operators exp($-i H t/\hbar $) and exp($- H /k T $), since now these operators
work no longer in the sense of exact  physical operations.}} Since the former 
deviations are not observed yet (for instance, as deviations from the optical 
theorem [3]), there should be a natural mechanism making them very small 
indeed. Let us emphasize that they would manifest themselves as a quantum 
effect, caused by a thermodynamic--like structure of a properly modified 
quantum dynamics. Such an effect is not present in the Einsteinian classical 
theory of gravitation.

 In this note we describe a thermodynamic--like model of quantum dynamics 
implying tiny deviations from the unitary time evolution as it is defined by 
the conventional state equation [4]

\begin{equation}
i\hbar{\frac{d\psi(t)}{dt}} = H\psi(t)\;,\;\; H^\dagger = H
\end{equation}

\ni and an initial condition for the state vector $\psi(t) $ (here, in some 
phenomenological situations, $ H $ may be time--dependent). The model is based 
on the correspondence (1) between time and absolute temperature, and so, in 
analogy with thermodynamics, it may be called "chronodynamics".

\vspace{0.3cm}

\ni {\bf 2. Equations of chronodynamics} 

\vspace{0.3cm}

 In the case of small deviations from the thermal equilibrium when $ T 
\rightarrow T + \delta T\rt $, the heat conductivity equation holds for $\delta
T\rt $. In a homogeneous matter medium, it takes the form

\begin{equation}
\left( \triangle - \frac{1}{\lambda_Q\,c}\frac{\partial}{\partial t}\right)\,
\delta T\rt = 0\; .
\end{equation}

\ni Here, $\delta T\rt \equiv 0 $ for the thermal equilibrium.

 By analogy, in the case of small deviations from the unitary time evolution 
({\it i.e.}, from quantum dynamical equilibrium) when the uniform time run 
would be a little deformed: $ t \rightarrow t + \delta t\rt $, a new conduct\-%
ivity equation should be valid for the inverse--time--deformation field

\begin{equation}
\varphi(\vec{r},t) \equiv \frac{1}{t + \delta t\rt} - \frac{1}{t} \simeq - 
\frac{\delta t\rt}{t^2}
\end{equation}

\ni (since the field $ -i\hbar\varphi\rt $ would be related to the field $ k
\delta T\rt $ through the corresp\-ond\-ence (1) ). In the vacuum, such an 
equation would have the form

\begin{equation}
\left( \triangle - \frac{1}{\lG c}\frac{\partial}{\partial t}\right)
\,\varphi\rt = 0 \; ,
\end{equation}

\ni where $ \lG > 0 $ would denote an unknown length--dimensional conductivity 
constant (in the vacuum). Here, $\delta t\rt \equiv 0 $ and so $\varphi\rt 
\equiv 0 $ for the unitary time evolution.

 Now, let us take into account the identity

\begin{eqnarray}
\left( \Box + \frac{1}{4\lG^2}\right) \left[\varphi\rt \exp{\frac{ct}{2\lG}} 
\right] & = & \exp{\frac{ct}{2\lG}}\left( \Box - \frac{1}{\lG c}
\frac{\partial}{\partial t}\right)\varphi\rt \nonumber \\  & \simeq & 
\exp{\frac{ct}{2\lG}} \left(\triangle - \frac{1}{\lG c} 
\frac{\partial}{\partial t}\right)\varphi\rt \; ,
\end{eqnarray}

\ni where the last step is valid if $(1/c^2)\partial^2 \varphi/\partial t^2 $
can be neglected nonrelativistically in comparison with $(1/\lG c)\,\partial
\varphi/\partial t $. Due to this identity, the conductivity equation (5) may 
be considered as a nonrelativistic approximation of tachyonic--type Klein---%
Gordon equation

\begin{equation}
\left( \Box + \frac{1}{4\lG^2}\right)\chi(x) = 0
\end{equation}

\ni for the new field

\begin{equation}
\chi(x) \equiv \varphi\rt \exp{\frac{ct}{2\lG}} \; .
\end{equation}

\ni Such a field equation for $ \chi(x) $ is relativistic in the sense of 
special relativity, if $ \chi(x) $ is a Lorentz covariant. Let us accept this 
equation and assume that $ \chi(x) $ is simply a Lorentz scalar. Then, in 
contrast to  $ \chi(x) $, the inverse--time--deformation field  $ \varphi\rt 
=\chi(x)\exp(-ct/2\lG) $, satisfying relativistically the equation

\begin{equation}
\left( \Box - \frac{1}{\lG c}\frac{\partial}{\partial t}\right)\varphi\rt = 0
\end{equation}

\ni in place of Eq. (5), is no Lorentz covariant. Of course, both fields $
\varphi\rt $ and $\chi(x) $ are real--parameter--valued, since time $ t $ is 
always a real parameter [4]. This implies that they cannot transport the 
quantum energy. It is important to observe that Eq. (7) allows for ultraluminal
plane--wave solutions $\exp (-ik\cdot x) $ with $ k_0 = \sqrt{\vec{k}^2 - 
1/4\lG^2} > 0 $, beside the damped--in--time plane--wave solutions with $ i 
k_0 = \sqrt{1/4\lG^2 - \vec{k}^2} > 0 $ (where $ k_0 = -i |k_0|$).

 Having accepted the homogeneous field equation (7), we come to a crucial point
in our argument: we must guess the form of matter sources which are to be 
inserted into this equation in order to describe the production of time--run 
deformations $\delta t\rt $ by time--evolving matter and, {\it vice versa}, 
their influence on the time evolution of matter. Here, the term "matter" means 
all physical particles, both fermions and bosons (including photons).

 The best guess we can offer is that the inhomogeneous field equation for the 
Lorentz scalar $\chi(x) \equiv \varphi\rt \exp(ct/2\lG)$ ought to have the form

\begin{equation}
\left( \Box + \frac{1}{4\lG^2}\right)\chi(x) = - g_\Gamma\lG\partial_\mu 
j^\mu(x)\; ,
\end{equation}

\ni where $ g_\Gamma > 0 $ is an unknown coupling constant and $\left(j^\mu(x)
\right) = \left( c\rho\rt\,,\;\vec{\jmath}\rt \right) $ describes a total 
average matter four--current. In the simplest case of pure quantum states, such
a current ought to be given by the spin--averaged expectation value

\begin{equation}
j^\mu(x) \equiv \langle\psi(t)|J^\mu(\vr)|\psi(t)\rangle_{\rm av}
\end{equation}

\ni with $\left(J^\mu(\vr) \right)$ denoting the operator of total particle 
four--current and $\psi(t) $ standing for the state vector of the considered
quantum system (here, $\psi(t) $ and $J^\mu(\vr) $ are presented in the 
Schr\"{o}dinger picture). Evidently, the state vector $\psi(t)$ in Eq. (11)
must satisfy a state equation, modified in comparison with the conventional 
state equation (2) valid in the case of unitary time evolution ({\it i.e.}, in 
the case of quantum dynamical equilibrium which holds when $\delta t\rt \equiv
0 $).{\footnote{In the general case of mixed quantum states, $ j^\mu(x) \equiv 
{\rm Tr}[ J^\mu(\vr)\rho(t)]$, where $\rho (t)$ is the system's density matrix
satisfying an extended Liouville---von Neumann equation, properly modified in 
comparison with its conventional form valid for the unitary time--evolution. 
Note that an extended form of this equation proposed in Ref. [5] is intended 
to modify the coventional quantum mechanics by introducing some decoherence 
effects at the microscopic scale, without simultaneously violating the proba\-%
bility conservation: $ d {\rm Tr}\rho (t)/dt = 0 $ (which is slightly violated 
in our model, {\it cf.} Eq. (20) and footnote $^4$ ).}} 

 The conjecture (10) means physically that time--run deformations are produced 
by local changes of the total average particle number and, {\it vice versa}, 
such local changes are influenced by time--run deformations.

 In order to complete the formal stucture of our model we must guess the 
modified state equation for $\psi(t) $. To this end let us observe that the 
correspondence (1) suggests for the first law of thermodynamics the modified 
form

\begin{equation}
dU = \delta W + \delta Q - i \delta \Gamma
\end{equation}

\ni with an imaginary term $-i \delta \Gamma $ added. Here, $-i \Gamma $ denotes
a new thermodynamic--like quantity being an analogue of heat Q, when $-i\hbar/t
$ takes over the role of $ kT $ according to Eq. (1). Let us call $\Gamma$ 
"energy width" transferred to the quantum system from the physical spacetime 
playing in our model the role of an unavoidable part of surroundings for any 
matter system. In analogy with the familiar heat law

\begin{equation}
Q \propto \int d^3\,\vr\,\rho\rt\,k\,\left[T+\delta T\rt - T\right]\; ,
\end{equation}

\ni we expect to have the formula

\begin{equation}
-i \Gamma \propto \int d^3\,\vr\,\rho\rt(-i\hbar)\left[\frac{1}{t + 
\delta\,t\rt} - \frac{1}{t} \right] \; ,
\end{equation}

\ni because $ -i \hbar/t $ corresponds to $ kT $. Then, for the unitary time 
evolution where $\delta\,t\rt \equiv 0 $, we get $ \Gamma \equiv 0 $.

 The modified first law of thermodynamics (12) leads in a natural way to the 
modified state equation of the form

\begin{equation}
i \hbar \frac{d\,\psi(t)}{dt} = \left(H - i{\bf 1}\Gamma\right)\,\psi(t)\;\; ,
\;\; H^\dagger = H \; ,
\end{equation}

\ni where {\bf 1} is the unit operator. Evidently, the new time--evolution 
generator $ H - i{\bf 1}\Gamma $ is (generally) a non--Hermitian operator 
implying nonunitary time evolution. We will accept Eq. (15) and assume that in 
the nonrelativistic approximation (with respect to matter) the energy width in
Eq. (15) can be presented as

\begin{equation}
\Gamma \equiv g_\Gamma\,\hbar\int\,d^3\vr\,\rho\rt\,\varphi\rt
\end{equation}

\ni in consistency with the formula (14), while its relativistically exact form
(with respect to matter) is

\begin{equation}
\Gamma \equiv g_\Gamma\,\hbar\int\,d^3\vr\,\sqrt{j_\mu(x)\,j^\mu(x)}\,
\varphi\rt\; .
\end{equation}

\ni Here, $g_\Gamma > 0 $ denotes the same coupling constant as that introduced
already in Eq. (10). It is expected to be very small in order to get tiny 
deviations from the unitary time evolution. Note from Eqs. (17) and (10) that 
$\Gamma = O(g_\Gamma^2) $. It is important to observe from Eqs. (15) and (17)
that the inverse--time--deformation field $\varphi\rt $, though it cannot 
transport quantum energy, can transfer to the quantum system the energy width.

 Since $\varphi\rt =\chi(x)\exp(-ct/2\lG)$ is no Lorentz covariant, the energy 
width $\Gamma $ as given in Eq. (17) is no pure time component of a four--%
vector, and so, while in Eq. (15), introduces necessarily some deviations from
the special relativity. They vanish (only) at $ t \rightarrow 0 $, what is well
defined if the Lorentz scalar $\chi(x) $ is regular at $ t = 0 $ or has there a
pole (for such a $\chi(x) $, the time--deformation field $\delta t\rt $, equal 
by Eq. (4) to $\delta t\rt = -\varphi\rt\,t^2\left[1 + \varphi\rt\,t \right]^{
-1}$, vanishes at $ t = 0 $). These deviations are caused by the factor $\exp (
-ct/2\lG) $ in $\varphi\rt $, so they may be avoided if in place of $\varphi
\rt $ the scalar $\chi (x)$ is used in Eqs. (16) and (17). However, such a 
choice for $\Gamma $ seems unnatural as long as the strict correspondence 
between Eqs. (13) and (14) has to be maintained. We will return to this 
potentially attractive choice at the end of Section 3.

 Thus, in our model, due to the factor $ \exp(-c t/2\lG) $ involved in the 
inverse--time--deformation field $ \varphi\rt $, the instant $ t = 0 $ is 
distinguished from all other instants in the evolution of Universe. This seems 
to suggest that time $ t $ in our model ought to be identified with the 
"cosmological time" $ t \geq 0 $ counted from the Big Bang as from its natural 
beginning at $ t = 0 $. In such a case, the analogy between the absolute 
temperature $ T \geq 0 $ and the inverse of time $ t\geq 0 $ really appeals to 
our imagination. For such reckoning of time we have $t = $ (running) age of 
Universe.

 Note from Eq. (15) that

\begin{equation}
\psi(t) = \psi^{(0)}(t) \exp\left[-\frac{1}{\hbar}\int_{t_0}^t dt'
\Gamma(t') \right] \; ,
\end{equation}

\ni where $\psi^{(0)}(t) $ satisfies the conventional state equation (2) 
describing the unitary time evolution. If the Hamiltonian $ H $ is time--%
independent, then

\begin{equation}
\psi^{(0)}(t) = \exp\left[-\frac{i}{\hbar} H (t -t_0)\right]\psi^{(0)}(t_0)\; .
\end{equation}

\ni We can see from Eq. (18) that the norm of $\psi(t) $ is equal to

\begin{equation}
\langle\psi(t)|\psi(t)\rangle^{1/2} = \exp\left[-\frac{1}{\hbar}\int_{t_0}^t
dt' \Gamma(t') \right] \; ,
\end{equation}

\ni so, generally, it changes (very slowly) in time, in contrast to $\langle
\psi^{(0)}(t)|\psi^{(0)}(t)\rangle^{1/2} = 1 $.{\footnote{In the general case 
of mixed quantum states, ${\rm Tr}\rho (t) = \sum_n \rho_n \exp[-(2/\hbar)
\int_{t_0}^t dt' \Gamma_n(t')] $ for $\rho (t) = \sum_n |\psi_n (t) \rangle 
\rho_n \langle \psi_n (t)|$ with $\sum_n \rho_n = 1 $ and $ H\psi_n (t_0) = 
E_n \psi_n (t_0)$. Thus, generally, it changes (very slowly) in time.}} The 
formal reason for this variation in time is that in our thermodynamic--like 
model of quantum dynamics the physical spacetime is not included into the 
quantum system as its part (in spite of mutual inter\-actions) [6], playing (in
virtue of these interactions) the role analogical to that of a thermostat in 
thermodynamics. In particular, during the unitary time evolution of a quantum 
system, the physical spacetime behaves formally as if it maintained in the 
quantum system the conventional uniform run of time $ t $, much like in the 
thermal equilibrium a thermostat keeps temperature $ T $ of a system uniform 
(and constant in time, though on cosmological scale the absolute temperature 
of Universe is also "running"). So, the term "chrono\-dynamics" for our theory 
seems to be justified on a profound level.

Concluding this Section, we can see that Eqs. (10) and (15) (together with (11)
and (17) ) form a mixed set of two coupled equations for the parameter--valued 
field $\chi(x)\equiv \varphi\rt\exp(ct/2\lG)$ describing time--run deformations
$\delta t\rt $ and, on the other hand, for the state vector $\psi(t) $ of 
time--evolving matter. Because of the bilinear form appearing in Eq. (11) this 
set of equations is (strictly speaking) nonlinear with respect to the state 
vector $\psi(t) $, what (slightly) violates the superposition principle for 
$\psi(t) $. This perturbs the fundam\-ental probabil\-ity inter\-pretation of 
$\psi(t)$.

 However, in the lowest--order perturbative approximation with respect to $ 
g_\Gamma $, where $\psi(t) $ in Eq. (11) is approximated by $\psi^{(0)}(t) $ 
satisfying the conventional state equation (2), the set of Eqs. (10) and (15)
becomes linear with respect to $\psi^{(1)}(t) $. In fact, in this approximation

\begin{equation}
\left( \Box + \frac{1}{4\lG^2}\right)\chi^{(1)}(x) = - g_\Gamma\lG
\partial_\mu j^{(0)\,\mu}(x)
\end{equation}

\ni with 

\begin{equation}
j^{(0)\,\mu}(x) \equiv \langle\psi^{(0)}(t)| J^\mu(\vr) |\psi^{(0)}(t)
\rangle_{\rm av}
\end{equation}

\ni and

\begin{equation}
i \hbar \frac{d\,\psi^{(1)}(t)}{dt} = \left(H - i{\bf 1}\Gamma^{(1)}\right)\,
\psi^{(1)}(t)\;\; ,\;\; H^\dagger = H \; ,
\end{equation}

\ni with

\begin{equation}
\Gamma^{(1)} \equiv g_\Gamma \hbar \int\,d^3\vr\sqrt{j^{(0)}_\mu (x) 
j^{(0)\,\mu} (x)}\,\varphi^{(1)}\rt = O(g^2_\Gamma)\; .
\end{equation}

\ni Thus, the state vector $\psi^{(1)}(t)$ still may be interpreted probabil\-%
istic\-ally in spite of a (very slow) variation in time of its norm

\begin{equation}
\langle\psi^{(1)}(t)|\psi^{(1)}(t)\rangle^{1/2} = \exp\left[-\frac{1}{\hbar}
\int_{t_0}^t dt' \Gamma^{(1)}(t') \right] \; ,
\end{equation}

\ni caused by the physical spacetime that is not included into the quantum 
system (in spite of mutual interactions) [6]. 

\vspace{0.3cm}

\ni {\bf 3. Consequences}

\vspace{0.3cm}

 There is a number of consequences of the hypothetic chronodynamics, caused by 
its departures from the conventional quantum theory. Most of them may consist 
of tiny (but {\it in spe} measurable) deviations from the unitary time 
evolution. Let us consider two examples

\vspace{0.1cm}
 
\ni {\it (i) Violation of optical theorem}
   
\vspace{0.1cm}

 The lowest--order perturbative $ S $ matrix in chronodynamics is related to 
the conventional $ S $ matrix through the formula

\begin{equation}
S^{(1)} = S^{(0)} \exp\left[-\frac{1}{\hbar}\int_{-\infty}^\infty dt 
\Gamma^{(1)}(t) \right] \; ,
\end{equation}

\ni where $\Gamma^{(1)}(t) = O(g_\Gamma^2) $ determines the lowest--order 
unitarity defect (here, $ S^{(0)\dagger} S^{(0)} = {\bf 1} = S^{(0)} 
S^{(0)\dagger}$). In Eq. (26) the interval $-\infty\,,\,\infty$ symbolizes a 
time interval very long in comparison with a short reaction time when $
\Gamma^{(1)}(t) \neq 0 $. Hence, in place of the conventional optical theorem

\begin{equation}
\sigma^{(0)}_{{\rm tot}\,i} = \frac{(2\pi)^4 \hbar^2}{v_i} \frac{1}{\pi} {\rm
Im}\, R^{(0)}_{ii}
\end{equation}

\ni we obtain

\begin{equation}
\sigma^{(1)}_{{\rm tot}\,i} = \frac{(2\pi)^4 \hbar^2}{v_i} \frac{1}{\pi} 
{\rm Im}\, R^{(1)}_{ii} \exp\left[-\frac{1}{\hbar}\int_{-\infty}^\infty dt 
\Gamma^{(1)}(t) \right] \; ,
\end{equation}

\ni where $ R^{(0)}_{f\,i}$ and $ R^{(1)}_{f\,i}$ are elements of the reaction 
matrix $ R^{(0)} $ and $ R^{(1)} $, respectively, while the labels $ i $ and $ 
f $ refer to particular asymptotic states of two colliding particles (averaged 
over particle spins). Here, in addition,

\begin{equation}
\frac{{\rm Im}R^{(1)}_{ii}}{{\rm Im}R^{(0)}_{ii}} = \exp\left[-\frac{1}{\hbar}
\int_{-\infty}^\infty dt \Gamma^{(1)}(t) \right] \; .
\end{equation}

 To get an idea of the possible magnitude of unknown coupling constant $ 
g_\Gamma $ and time--deformation conductivity constant $ \lG $ we pass to the 
next example.

\vspace{0.1cm}
 
\ni {\it (ii) Variation of average particle number}
   
\vspace{0.1cm}

 Consider a sample of gas consisting of some identical nonrelativistic part\-%
icles, {\it e.g.} electrons or hydrogen atoms, situated in proximity of a 
pointlike target $ \vr_{\rm S} $ of a big accelerator producing $1/\tau $ 
particles of all sorts per unit of time. Then, according to Eq. (10), the 
accelerator during its stationary run excites time--run deformations $\delta
t^{\rm ex}\rt $ described by the static field of the form

\begin{equation}
\chi^{\rm ex}(\vr) = \frac{g_\Gamma\lG}{\tau} \frac{\cos(|\vr - \vr_{\rm S}|/2
\lG)}{|\vr - \vr_{\rm S}|} \sum_{l\,m_l} c_{l\,m_l}Y_{l\,m_l}(\theta,\phi)\;,
\end{equation}

\ni where $\sum_{l\,m_l} c_{l\,m_l}Y_{l\,m_l}(0,0) = 1 $ and $\theta\,,\;\phi $
are spherical angles for $\vr - \vr_{\rm S} $. This field satisfies the 
equation $ (\triangle + 1/4\lG^2)\,\chi^{\rm ex}(\vr) = = -4\pi g_\Gamma\lG 
{\rm div} \vec{\jmath}(\vr) $, where 

\begin{equation}
{\rm div}\vec{\jmath}(\vr) = \frac{1}{\tau}\left[ \delta^3(\vr - \vr_{\rm S}) 
+ \frac{\cos(|\vr - \vr_{\rm S}|/2\lG)}{|\vr - \vr_{\rm S}|^3} 
\sum_{l\,m_l} c_{l\,m_l}\frac{l(l+1)}{4\pi}\,Y_{l\,m_l}(\theta,\phi)\right]
\end{equation}

\ni gives a mathematical model of the particle current $\vec{\jmath}(\vr)$ 
produced on the target $\vr_{\rm S} $. The field $ \chi^{\rm ex}(\vec{r}) $, 
playing the role of an external field for our sample of probe particles, 
modifies the conventional wave function $ \psi^{(0)}(\vec{r}_{\rm P},t) = 
\psi^{(0)}(\vec{r}_{\rm P})\exp\left[-({i}/{\hbar})\,E\,(t - t_{0})\right] $ of
any particle in the sample, leading in virtue of Eq. (18) to the following 
lowest--order perturbed wave function:

\begin{equation}
\psi^{(1)}(\vec{r}_{\rm P},t) = \psi^{(0)}(\vec{r}_{\rm P})\exp\left[
-\frac{i}{\hbar}\,E\,(t - t_{0})\right]\exp\left[ - \frac{1}{\hbar}
\int_{t_{0}}^{t} dt'\,\Gamma^{(1)}(t')\right]\;.
\end{equation}

\ni Here, by Eq. (16)

\begin{equation}
\Gamma^{(1)}(t) \equiv  g_\Gamma \hbar \int_{V}\! d^3\vec{r} \rho^{(0)}(\vr)\,
\chi^{\rm ex}(\vec{r}) \,\exp\left(-\frac{c\,t}{2\lG}\right) = O(g_\Gamma^2)
\end{equation}

\ni and so 

\begin{equation}
\int_{t_0}^t dt' \Gamma^{(1)}(t') = \frac{2g_\Gamma\lG\hbar}{c} \int_{V}\! 
d^3\vec{r} \rho^{(0)}(\vr)\,\chi^{\rm ex}(\vec{r})\,\exp\left(-\frac{c\,t_0}{2
\lG}\right) \left\{1 - \exp\left[-\frac{c(t - t_0)}{2\lG} \right] \right\}\; .
\end{equation}

\ni In our argument, the stationary run of accelerator is switched on at the 
moment $t_0 $ and still lasts at the later moment $ t $.

 In this example, when using Eq. (11) with the operator of particle density of 
the form $ (1/c)J^0(\vr) = \delta^3(\vr - \vr_{\rm P}) $, we get

\begin{equation}
\rho^{(0)}(\vec{r}) = \int_{V}\! d^3\vec{r}_{\rm P}\,|\psi^{(0)}
(\vec{r}_{\rm P})|^2\,\delta^3(\vec{r} - \vec{r}_{\rm P}) = \frac{1}{V}\;\;\;
{\rm for}\;\;\; \vr \in V
\end{equation}

\ni if we put $\psi^{(0)}(\vec{r}_{\rm P}) = (1/\sqrt{V})\exp(i\vec{k}\cdot
\vr_{\rm P}) $. In a similar way, we have

\begin{equation}
\rho^{(1)}(\vec{r}) = \frac{1}{V}\exp\left[-\frac{2}{\hbar}\int_{t_0}^t dt'\, 
\Gamma^{(1)}(t') \right]\;\;\;{\rm for}\;\;\; \vr \in V\; .
\end{equation}

\ni Hence, the lowest--order perturbed average number of particles contained in
the sample is given by

\begin{equation}
N^{(1)}(t) = N^{(0)}\exp\left[-\frac{2}{\hbar} N^{(0)} \int_{t_0}^t dt' 
\Gamma^{(1)}(t') \right]\; ,
\end{equation}

\ni where, in particular for gas of hydrogen atoms, $ N^{(0)} = N^{(1)}(t_0) $ 
is equal (in normal conditions) to the Loschmidt number 2.69$\times 10^{19}
{\rm cm}^{-3}$ multiplied by the volume $ V $. Thus, such an average number of 
particles changes in time during the stationary run of the accelerator. Of 
course, this variation is expected to be very slow. For a sample of charged 
particles, {\it e.g.} electrons, this conclusion implies unavoidably a simul\-%
taneous variation in time of the total average charge.

 Using Eqs. (33), (35) and (30) we can write in Eq. (37)

\begin{eqnarray}
\int_{t_0}^t dt' \Gamma^{(1)}(t') & = & \frac{2g_\Gamma\lG\hbar}{c}\frac{1}{V}
\int_{V}\! d^3\vec{r}\,\chi^{\rm ex}(\vec{r})\,\exp\left(-\frac{c\,t_0}{2\lG}
\right) \left\{1 - \exp\left[-\frac{c(t - t_0)}{2\lG} \right] \right\} 
\nonumber \\ & \simeq & \frac{2g_\Gamma^2\lG^2\hbar}{c\,\tau}\,
\frac{\cos(d/2\lG)}{d}\,\exp\left(-\frac{c\,t_0}{2\lG}\right) \left\{1 - 
\exp\left[-\frac{c(t - t_0)}{2\lG} \right] \right\} \; ,
\end{eqnarray}

\ni where $ d $ denotes an average distance of the sample from the pointlike 
target of accelerator, while the average value of angular part of 
$\chi^{\rm ex}(\vr) $ over the sample is put equal to 1. If $c(t - t_0)/2\lG 
\ll 1 $ or $c(t - t_0)/2\lG \gg 1 $, this estimation gives

\begin{eqnarray}
\int_{t_0}^t dt'\,\Gamma^{(1)}(t') & \simeq & \frac{g_\Gamma^2\,\lG\,\hbar}
{\tau}\,\frac{\cos(d/2\lG)}{d}\,\exp\left(-\frac{c\,t_0}{2\lG}\right) (t - t_0)
\nonumber \\ & \simeq & \frac{g_\Gamma^2\,\lG\,\hbar}{\tau\,d} \exp\left(
-\frac{c\,t_0}{2\lG}\right) (t - t_0) > 0 
\end{eqnarray}

\ni or

\begin{eqnarray}
\int_{t_0}^t dt' \Gamma^{(1)}(t') & \simeq & \frac{2g_\Gamma^2\,\lG^2\,\hbar}
{c\,\tau}\,\frac{\cos(d/2\lG)}{d}\,\exp\left(-\frac{c\,t_0}{2\lG}\right) 
\nonumber \\ & \simeq & \frac{2g_\Gamma^2\,\lG^2\,\hbar}{c\,\tau\,d}\,
\exp\left(-\frac{c\,t_0}{2\lG}\right) \simeq 0 \; ,
\end{eqnarray}

\ni respectively. Here, the last step is valid if, in addition, $ d/2\lG \ll 1 
$ or $ d/2\lG \gg 1 $ (in the second case, because of $[\cos(d/2\lG) - 1]
(d/2\lG)^{-1} \rightarrow - \pi\,d\delta(d) = 0 $ for $\lG \rightarrow 0 $). 
We can see from Eqs. (37) and (39) or (40) that in the first case (where $ c
(t - t_0)/2\lG \ll 1 $ and $ d/2\lG \ll 1 $) the number $N^{(1)}(t) $ decreases
(very slowly) in time, or in the second case (where $c(t - t_0)/2\lG \gg 1$ and
$ d/2\lG \gg 1$) its variation in time is completely negligible, what excludes 
in this case the positive observation of possible nonunitarity effects.

 In order to get an idea of the possible magnitude of unknown constants $ 
g_\Gamma $ and $\lG $ in the decrement factor in Eq. (37), let us consider our 
first case and put: $ N^{(0)} = 2.69\times 10^{19} V {\rm cm}^{-3} $, the 
accelerator total production rate $ 1/\tau \sim  10^{8}\,{\rm sec}^{-1}\;,\;
V/d \sim 100\,{\rm cm}^2 $ and $ t - t_0 \sim 1 $ month $\sim 10^6 $ sec.
Then, making use of Eq. (39), we obtain

\begin{equation}
p^{(1)} \equiv \frac{2}{\hbar}N^{(0)}\int_{t_0}^t dt'\,\Gamma^{(1)}(t') \sim
10^{36}g_\Gamma^2\lG\exp \left(-\frac{c\,t_0}{2\lG}\right){\rm cm}^{-1}\; .
\end{equation}

\ni Hence, requiring for the decrement factor in Eq. (37) the lower bound lying
(for instance) in the range

\begin{equation}
\exp(-p^{(1)}) \stackrel{>}{\sim} 0.9900\;\;{\rm to}\;\; 0.9999
\end{equation}

\ni we get the estimation
 
\begin{equation}
g_\Gamma^2\lG \exp\left(-\frac{c\,t_0}{2\lG}\right) \stackrel{<}{\sim}(10^{-38}
\;\;{\rm to}\;\;10^{-40}){\rm cm}
\end{equation}

\ni (here, the exponent is practically fixed for time $ t_0 $ counted from the 
Big Bang).

 Finally, in consistency with our first case, let us consider for the length--%
dimensional conductivity constant $\lG $ the extreme cosmological option: $\lG
\sim c\times {\rm age\;of\;Universe} \sim 10^{28}\,$cm (if age of Universe 
$\sim 1.5\times 10^{10}\,$yr). Then, under our conjecture that time $ t $ ought
to be counted from the Big Bang, we have $ t_0 \sim t = \;$age of Universe 
and $\exp(-c t_0/2\lG) \sim \exp(-1/2) = 0.6065 $, and so $ g_\Gamma^2 
\stackrel{<}{\sim} 10^{-66}$ to $10^{-68}$ from Eq. (43).

 Concluding this Section, we can say that, at present, it seems difficult to 
estimate the magnitude of $ g_\Gamma $ and $\lG $, unless we decide to accept 
for $\lG $ the above extreme option (or another cosmological option postulating
also a big $\lG $, favorable from the viewpoint of experimental discovery
potential).

 However, the situation changes essentially, if in Eqs. (16) and (17) (defining
the energy width $\Gamma $) the scalar $\chi(x)\equiv \varphi\rt\exp(ct/2\lG)$
is used in place of $\varphi\rt$ (what in Section 2 was considered an unnatural
choice in view of the strict correspondence between Eqs. (13) and (14), though 
this choice is potentially attractive, giving no deviations from the special 
relativity). Now, for such a choice, the factor $\exp(-ct/2\lG) $ disappears 
from Eq. (33) and, in consequence, Eq. (34) is replaced by the formula

\begin{equation}
\int_{t_0}^t dt' \Gamma^{(1)}(t') = g_\Gamma\hbar \int_{V}\! d^3\vec{r} 
\rho^{(0)}(\vr)\,\chi^{\rm ex}(\vec{r})(t - t_0)\; .
\end{equation}

\ni Then, instead of Eq. (39) or (40), the relation

\begin{equation}
\int_{t_0}^t dt'\,\Gamma^{(1)}(t') \simeq \frac{g_\Gamma^2\lG\hbar}{\tau\,d}
(t - t_0) > 0
\end{equation}

\ni holds for any $c(t - t_0)/2\lG $ in both the cases $d/2\lG \ll 1$ and $ d/
2\lG \gg 1$. Hence, with the use of lower bound (42), the estimation (43), but 
now without the factor $\exp(-ct_0/2\lG)$ on its lhs, follows.

 Thus, for the extreme cosmological option $\lG \sim c \times $ age of Universe
$\sim 10^{28}\,$cm, consistent with our first case, the previous estimate $ 
g_\Gamma^2 \stackrel{<}{\sim} 10^{-66}$ to $10^{-68} $ does not change in the 
new situation, since for this option the irrelevant factor $\exp(-ct_0/2\lG) 
\sim \exp(-1/2) \sim 1 $ (with $t_0 \sim t = $ age of Universe) was used 
previously in Eq. (43).

 But, in this new situation, even the extreme microscopic option $\lG \sim $ 
Planck length $ = \hbar/M_{\rm PL}\,c \sim 10^{-33}\,$cm, consistent with our 
second case, is experimentally not excluded {\it a priori}, because the factor 
$\exp(-ct/2\lG)$ (being as small as $\exp(-61) $ for this option) is now absent
from Eq. (33). For such an option the estimate $ g_\Gamma^2 \stackrel{<}{\sim}
10^{-5} $ to $ 10^{-7} $ follows from Eq. (43), if the factor $-\exp(ct_0/2\lG)
$ on its lhs is omitted.

 Note that $\Gamma $ as given in Eq. (17) is replaced now by the choice

\begin{equation}
\Gamma \equiv g_\Gamma \hbar \int_V d^3 \vr \sqrt{j_\mu(x) j^\mu(x)}\, \chi(x) 
= g_\Gamma \hbar \exp \frac{ct}{2\lG} \int_V d^3 \vr \sqrt{j_\mu(x) j^\mu(x)}
\,\varphi\rt
\end{equation}

\ni which, in the nonrelativistic approximation (with respect to matter), is 
also consistent with Eq. (14), but only within a narrow time interval $0 \leq t
- t_0 \ll 2\lG/c $ (for any $ t_0 > 0$). In the case of a cosmological option,
extremally of $\lG \sim 10^{28}\,$cm, such a time interval is not necessarily 
so narrow.

 Since $\Gamma $ as given in Eq. (46) is pure time component of a Lorentz 
four--vector, it is consistent with the special relativity, while inserted into
Eq. (15). We prefer, therefore, this choice, though it is not so close to Eq.
(14) as the choice (17).

\vfill\eject

~~~~
\vspace{0.5cm}

{\bf References}

\vspace{1.0cm}

{\everypar={\hangindent=0.5truecm}
\parindent=0pt\frenchspacing

~1.~Cf. {\it e.g.}~A.L.~Fetter and J.D.~Walecka, {\it Quantum theory of 
many--particle systems}, McGraw--Hill (1971).

~2.~W.~Kr\'{o}likowski, {\it Acta Phys. Pol.} {\bf B 24}, 1903 (1993); {\it 
Nuovo Cimento} {\bf 107 A}, 1759 (1994).

~3.~P.H. Eberhard {\it et al, Phys. Lett.} {\bf 53 B}, 121 (1974).

~4.~P.A.M.~Dirac, {\it The principles of quantum mechanics}, 4th ed., Oxford
University Press (1959).

~5. J. Ellis, J.S. Hagelin, D.V. Nanopoulos and M. Srednicki, {\it Nucl. Phys.}
{\bf B 241}, 381 (1984); J. Ellis, J.L. Lopez, N.E. Mavromatos and D.V. 
Nanopoulos, {\it Phys. Rev.} {\bf D 53}, 3846 (1996).

~6. For a general formalism of projecting out an interacting subsystem from a 
larger quantum system {\it cf.}~W.~Kr\'{o}likowski and J.~Rzewuski, {\it Nuovo 
Cimento }{\bf 25 B}, 739 (1975), and references therein. Such a projection 
leads generally to a nonzero energy--width operator for the remaining quantum 
subsystem, in particular, to a decay--width operator. In the case of chrono\-%
dynamics, the remaining quantum subsystem is the whole matter system containing
particles of all sorts, while the subsystem projected out is the physical 
spacetime (the possibility of constructing a larger quantum system including 
the spacetime as its subsystem is anticipated in this argument: this inclusion 
should lead to quantum gravity).
 
\end{document}